\documentclass[12pt]{article}
\begin{document}
\pagenumbering{roman}
\def\Universita{Universit\`a}
\def\Barcelo{Barcel\'o}
\title{\bf Einstein gravity as an emergent phenomenon?}
\author{%
Carlos {\Barcelo} and Matt Visser\\[2mm]
{\small \it 
Physics Department, Washington University, 
Saint~Louis, Missouri 63130-4899, USA}
\\[12pt]
Stefano Liberati\\[2mm]
{\small \it 
Physics Department, University of Maryland, 
College Park, MD 20742-4111, USA}
\\[12pt]
} 
\date{{\small 22 March 2001; \LaTeX-ed \today}\\
{\small \it Paper awarded of an ``honorable mention'' in
    the Annual Competition\\ of the Gravity Research Foundation for the
    year 2001.}}
\maketitle
\vfill
\hrule
\bigskip
\centerline{\underline{E-mail:} {\sf carlos@hbar.wustl.edu}}
\centerline{\underline{E-mail:} {\sf visser@kiwi.wustl.edu}}
\centerline{\underline{E-mail:} {\sf liberati@physics.umd.edu}}
\bigskip
\centerline{\underline{Homepage:} {\sf http://www.physics.wustl.edu/\~{}carlos}}
\centerline{\underline{Homepage:} {\sf http://www.physics.wustl.edu/\~{}visser}}
\centerline{\underline{Homepage:} {\sf http://www2.physics.umd.edu/\~{}liberati}}

\bigskip
\bigskip
\hrule
\clearpage
\markright{Einstein gravity as an emergent phenomenon?\hfil }
\pagestyle{myheadings}

\null
\vskip 2cm

\begin{abstract}


\bigskip

In this essay we marshal evidence suggesting that Einstein gravity may
be an emergent phenomenon, one that is not ``fundamental'' but rather
is an almost automatic low-energy long-distance consequence of a wide
class of theories.  Specifically, the emergence of a curved spacetime
``effective Lorentzian geometry'' is a common generic result of
linearizing a classical scalar field theory around some non-trivial
background.  This explains why so many different ``analog models'' of
general relativity have recently been developed based on condensed
matter physics; there is something more fundamental going on.  Upon
quantizing the linearized fluctuations around this background
geometry, the one-loop effective action is guaranteed to contain a
term proportional to the Einstein--Hilbert action of general
relativity, suggesting that while classical physics is responsible for
generating an ``effective geometry'', quantum physics can be argued to
induce an ``effective dynamics''. This physical picture suggests that
Einstein gravity is an emergent low-energy long-distance phenomenon
that is insensitive to the details of the high-energy short-distance
physics.

\vspace*{5mm}
\end{abstract}

\vfill

\clearpage

\def\half{{1\over2}}
\def\L{{\mathcal L}}
\def\S{{\mathcal S}}
\def\d{{\mathrm{d}}}
\def\etal{{\emph{et al}}}
\def\det{{\mathrm{det}}}
\def\tr{{\mathrm{tr}}}
\def\ie{{\emph{i.e.}}}
\def\aka{{\emph{aka}}}
\def\Choose#1#2{{#1 \choose #2}}
\def\Eotvos{{E\"otv\"os}}
\def\HRULE{{\bigskip\hrule\bigskip}}



\section{Analog gravity}
\setcounter{page}{1}
\pagenumbering{arabic}

There is a risk that all current attempts at ``quantizing gravity''
are condemned to failure because they have been starting from
fundamentally flawed premise, and that in reality there are no
fundamental gravitational degrees of freedom to quantize --- it is
possible that Einstein gravity is an emergent phenomenon, in the same
sense that fluid dynamics emerges from molecular physics as a
low-momentum long-distance approximation. In this essay we will take a
careful look at this idea, and highlight some of the possibilities,
problems, and opportunities that such a situation entails.

We were led to these notions via current research on analog models of
general relativity~\cite{Workshop}. Because of the extreme difficulty
(and inadvisability) of working with intense gravitational fields in a
laboratory setting, interest has now turned to investigating the
possibility of {\emph{simulating}} aspects of general relativity ---
though it is not {\emph{a priori}} expected that all features of
Einstein gravity can successfully be carried over to the analog
models.  Numerous rather different physical systems have now been
seen to be useful for developing analog models of general relativity.
A literature search as of March 2001 finds well over a hundred
scientific articles devoted to one or another aspect of analog gravity
and effective metric techniques. The sheer number of different
physical situations lending themselves to an ``effective metric''
description strongly suggests that there is something deep and
fundamental going on.

Typically these are models {\em of} general relativity, in the sense
that they provide an effective metric and so generate the basic
kinematical background in which general relativity resides; in the
absence of any dynamics for that effective metric we cannot really
speak about these systems as models {\em for} general relativity.
However, as we will discuss more fully bellow, quantum effects in
these analog models might provide of some sort of dynamics resembling
general relativity.

Remember that for mechanical systems with a finite number of degrees
of freedom small oscillations can always be resolved into normal
modes: a finite collection of uncoupled harmonic oscillators.  For a
classical field theory you would also expect similar behaviour: small
deviations from a background solution of the field equations will be
resolved into travelling waves; then these travelling waves can be
viewed as an infinite collection of harmonic oscillators, or a finite
number if the field theory is truncated in the infra-red and
ultra-violet, to which you can then apply a normal mode analysis. The
physically interesting question is whether this normal mode analysis
for field theories can then be reinterpreted in a ``geometrically
clean'' way in terms of some ``effective metric'' and ``effective
geometry''. In many cases the answer is definitely yes: Linearization
of a Lagrangian-based dynamics, or linearization of any hyperbolic
second-order PDE, will automatically lead to an effective Lorentzian
geometry that governs the propagation of the fluctuations. [The most
general situation (multiple scalar fields, or a multi-component vector
or tensor) is quite algebraically messy --- details of that situation
will be deferred for now.]

Once the notion of a derived ``effective metric'' has been
established, we can certainly consider the effect of quantizing the
linearized fluctuations. At one loop the quantum effective action will
contain a term proportional to the Einstein--Hilbert action --- this
is a key portion of Sakharov's ``induced gravity''
idea~\cite{Sakharov}. In the closing segment of the essay we argue
that the occurrence of not just an ``effective metric'', but also an
``effective geometrodynamics'' closely related to Einstein gravity, is
a largely unavoidable feature of the linearization and quantization
process.

\section{Effective metric}

Suppose we have a single scalar field $\phi$ whose dynamics is
governed by some first-order Lagrangian $\L(\partial_\mu\phi, \phi)$.
(By ``first-order'' we mean that the Lagrangian is some arbitrary
function of the field and its first derivatives.) We want to consider
linearized fluctuations around some background solution $\phi_0(t,\vec
x)$ of the equations of motion, and to this end we write
\begin{equation}
\phi(t,\vec x) = \phi_0(t,\vec x) + \epsilon  \phi_1(t,\vec x) + O(\epsilon^2).
\end{equation}
Linearizing the Euler--Lagrange equations results in a second-order
differential equation with position-dependent coefficients (these
coefficients all being implicit functions of the background field
$\phi_0$). Following an analysis developed for acoustic geometries
(Unruh~\cite{Unruh}, Visser {\etal}~\cite{Visser}), which also applies
to this much more general situation, this can be given a nice clean
geometrical interpretation in terms of a d'Alembertian wave equation
--- provided we \emph{define} the effective spacetime metric by
\begin{equation}
\sqrt{-g} \; g^{\mu\nu} \equiv  f^{\mu\nu} \equiv 
\left.\left\{
{\partial^2 \L\over\partial(\partial_\mu \phi) \; \partial(\partial_\nu \phi)} 
\right\}\right|_{\phi_0},
\end{equation}
that is,
\begin{equation}
g_{\mu\nu}(\phi_0) =    
\left.
\left( - 
\det\left\{
{\partial^2 \L\over\partial(\partial_\mu \phi) \; \partial(\partial_\nu \phi)} 
\right\}
\right)^{1/(d-1)}
\right|_{\phi_0}
 \; \; 
\left.
\left\{
{\partial^2 \L\over\partial(\partial_\mu \phi) \; \partial(\partial_\nu \phi)} 
\right\}^{-1}
\right|_{\phi_0}.
\end{equation}
The equation of motion for the linearized fluctuations can then be
written in the geometrical form
\begin{equation}
\label{E:geometrical}
\left[\Delta(g(\phi_0)) - V(\phi_0)\right] \phi_1 = 0,
\label{laeone}
\end{equation}
where $\Delta$ is the d'Alembertian operator associated with the
effective metric $g(\phi_0)$, and $V(\phi_0)$ is a
background-field-dependent potential.  It is important to realise just
how general the result is: it works for {\em any} Lagrangian depending
only on a single scalar field and its first derivatives. The
linearized PDE will be {\emph{hyperbolic}} (and so the linearized
equations will have wave-like solutions) if and only if the effective
metric $g_{\mu\nu}$ has Lorentzian signature
$\pm[-,(+)^d]$~\cite{Courant}.  Note that $d=1$ space dimensions is
special, and the present formulation does not work unless
$\det(f^{\mu\nu})=1$. This observation can be traced back to the
conformal covariance of the Laplacian in $1+1$ dimensions, and implies
(perhaps ironically) that the only time the procedure risks failure is
when considering a field theory defined on the world sheet of a
string-like object.

Indeed, even if you do not have a Lagrangian, it is still possible to
extract an ``effective metric'' for a system with one degree of
freedom. (More precisely, we can define a conformal class of effective
metrics.  The analysis is not as geometrically ``clean''.)  While
several of the technical details are different from the
Lagrangian-based analysis, the basic flavor is the same: The key point
is that hyperbolicity of the linearized PDE is defined in terms of the
presence of a matrix of indefinite signature
$\pm[-,(+)^d]$~\cite{Courant}. This matrix is enough to define a
conformal class of Lorentzian metrics, and picking the ``right''
member of the conformal class is largely a matter of taste --- do
whatever makes the ``geometrized'' equation look cleanest.

\section{Effective dynamics}

At this stage we have derived the existence of a background metric
$g_{\mu\nu}(\phi_0)$ and linearized fluctuations governed by the
equation (\ref{laeone}).  We shall now try to see how and to what
extent it is possible to define a {\emph{dynamics}} for this metric. Of
course it would be not particularly useful to search for such dynamics
in the equations of motion of the ``fundamental system'' [those
derived from ${\cal L}(\phi,\partial\phi)$]. It is from this
perspective obvious that the dynamical equations of the effective
metric should also be regarded as an ``emergent'' phenomenon. The idea
of ``induced gravity'', proposed several years ago by Andrei
Sakharov~\cite{Sakharov}, provides a natural framework for such an
emergence of an effective geometrodynamics. Along these lines we shall
now try to derive Einstein-like equations from the one-loop action of
the $\phi$ field.

Expand the field $\phi$ into a background $\phi_b$, which does not
necessarily satisfy any classical equation of motion, and a quantum
fluctuation $\phi_q$ so that $\phi = \phi_b + \phi_q$. Integrate out
the quantum fluctuations; then at one loop
\begin{equation}
\Gamma[g(\phi_b),\phi_b] = S[\phi_b] + 
{1\over2} \;\hbar \; {\rm tr}\ln \left[\Delta(g(\phi_b)) - V(\phi_b)\right]
+ O(\hbar^2).
\end{equation}
Here the determinant of the differential operator may be defined in
terms of zeta functions or heat kernel expansions~\cite{Blau}. 
Note also that the effective action depends on the background field in
two ways: {\emph{explicitly}} through $\phi_b$, and
{\emph{implicitly}} through $g(\phi_b)$. The key point 
is that defining the determinant requires both regularization and
renormalization, and that doing so introduces counterterms
proportional to the first $d/2$ Seeley-DeWitt
coefficients~\cite{Blau}.  The zeroth Seeley--DeWitt coefficient $a_0$
induces a cosmological constant, while $a_1$ induces an
Einstein--Hilbert term, and there are additional terms proportional to
$a_2$.  All in all:
\begin{eqnarray}
\Gamma[g(\phi_b),\phi_b] &=& S[\phi_b] + \hbar \int \sqrt{-g}\; \kappa \left[ -2
\Lambda + R(g)-6V(\phi_b)\right] \d^{d+1}x 
\nonumber\\
&&
+ \hbar X[g(\phi_b),\phi_b] + O(\hbar^2).
\end{eqnarray}
Here $X[g(\phi_b),\phi_b]$ denotes all other finite contributions to
the renormalized one-loop effective action. It is the automatic
emergence of the Einstein--Hilbert action as part of the one-loop
effective action that is the salient point.  Note that our approach is
not identical to Sakharov's idea --- in his proposal the metric was
put in by fiat, but without any intrinsic dynamics; all the dynamics
was generated via one loop quantum effects. In our proposal the very
existence of the effective metric itself is an emergent phenomenon. In
Sakharov's approach the metric was free to be varied at will, leading
precisely to the Einstein equations (plus quantum corrections); in our
approach the metric is not a free variable and the equations of motion
will be a little trickier.

The quantum equations of motion are defined in the usual way by
varying $\Gamma[\phi_b]$ with respect to the background $\phi_b$. It
is important to remember that the metric is a function of the
background field so that it does not make sense to vary the metric
independently --- we must always evaluate variations using the chain
rule. 
\begin{equation}
{\delta\Gamma[g(\phi_b),\phi_b]\over\delta\phi_b(x)} 
\equiv
\left.{\delta\Gamma[g(\phi_b),\phi_b]\over\delta\phi_b(x)}\right|_{g_b}
+ \left.{\delta\Gamma[g(\phi_0),\phi_0]\over\delta
g_{\mu\nu}}\right|_{\phi_b} {\delta
g_{\mu\nu}(\phi_b)\over\delta\phi_b(x)}.
\end{equation}
In this way we see that the equations of motion have a part coming
from the variation with respect to the background field plus a part
proportional to the variation with respect to the metric.  It is this
second part which provides an Einstein-like dynamics.  The presence of
the terms generated by variation with respect to $\phi_b$ leads to the
interesting conclusion that one needs to assume, in order to get a
dynamics as close as possible to that of Einstein, that these terms
satisfy a special constraint: there should exist a functional $Y[g]$,
depending only on the effective metric, such that
\begin{eqnarray}
\left\{
{\delta S[\phi_b]\over\delta \phi_b } 
- 6 \hbar \kappa {\delta \int \sqrt{-g} \; V(\phi_b)\over\delta \phi_b }
+ \hbar
\left.{\delta X[g(\phi_b),\phi_b]\over\delta \phi_b }\right|_{g_b}
\right\}   
&=& \hbar {\delta Y[g]\over\delta g_{\mu\nu}} 
  {\delta g_{\mu\nu}(\phi_b)\over\delta\phi_b(x)}
\nonumber\\
&& + O(\hbar^2).
\end{eqnarray}
In this case the background geometry decouples from the effective
metric and we have
\begin{equation}
\left[ 
\kappa \left(G^{\mu\nu}(g) + \Lambda g^{\mu\nu}\right) + 
{1\over\sqrt{g}}
{\delta \{X[g(\phi_b)]+Y[g(\phi_b)]\}\over\delta g_{\mu\nu} }
\right] \; 
{\delta g_{\mu\nu}(\phi_b)\over\delta\phi_b(x)} = O(\hbar).
\end{equation}
The ${\delta X[g(\phi_b)]/\delta g}$ term denotes the type of
``curvature squared'' correction to the Einstein equations that is
commonly encountered in string theory (indeed in almost any candidate
theory for quantum gravity), and also in the usual implementation of
Sakharov's approach.  However it must be emphasised that because of
the contraction with the ${\delta g_{\mu\nu}(\phi_b)/\delta\phi_b(x)}$
these are not the usual Einstein equations, though they are certainly
implied by the (curvature enhanced) Einstein equations. It is in this
sense that we can begin to see the structure of Einstein gravity
emerging from this field-theoretic normal mode analysis.

\section{Prospects}

In this essay we have argued that the emergence of an ``effective
metric'', in the sense that this notion is used in the so-called
``analog models'' of general relativity, is a rather generic feature
of the linearization process. While the existence of an effective
metric by itself does not allow you to simulate all of Einstein
gravity, it allows one to do quite enough to be really significant ---
in particular it seems that the existence of an effective Lorentzian
metric is really all that is in principle needed to obtain simulations
of the Hawking radiation effect~\cite{Unruh,without-entropy}.

By invoking one-loop quantum effects, we have argued that something
akin to Sakharov's induced gravity scenario is operative: in
particular we can generically argue that there is a term in the
quantum effective action proportional to the Einstein--Hilbert
action. However because of the technical assumption that the effective
metric depends on the background only via the scalar field $\phi_0(x)$
we have not been able to reproduce full Einstein gravity, though
certainly have some extremely suggestive results along this line.

The major steps that are needed to extend this idea to a full fledged
theory are
\begin{enumerate}
\item
The question of what happens when many fields are present in the
problem: the major piece of additional physics is the possible
presence of birefringence, or more generally ``multi-refringence'',
with different normal modes possibly reacting to different
metrics. The {\Eotvos} experiment [the observational universality of
free fall to extremely high accuracy] indicates that all the physical
fields coupling to ordinary bulk matter ``see'' to high precision the
{\emph{same}} metric, allowing us to formulate the Einstein
Equivalence principle and speak of {\emph{the}} metric of spacetime.
\item 
Whether the addition of extra fields helps one to obtain a better
approximation to full Einstein gravity --- this because you would get
one equation of motion per background field, so with six or more
fields you would expect to be able to explore the full algebraic
structure of the metric.  So adding extra fields, which is technically
a hindrance in the kinematical part of the program (developing the
effective metric formalism), should in compensation allow one to more
closely approach the dynamics of Einstein gravity.
\end{enumerate}

In summary: The full generality of the situations under which
effective metrics are encountered is truly remarkable, and the extent
to which the resulting analog models seem able to reproduce key
aspects of Einstein gravity is even more remarkable. The physics of
these systems is fascinating, and the potential for laboratory
investigation of models close to (but not necessarily identical to)
Einstein gravity is extremely encouraging.

Our interpretation of these results is that they provide suggestive
evidence that what we call Einstein gravity (general relativity) is an
almost automatic low-energy consequence of almost any well behaved
quantum field theory: the occurrence of an effective metric is almost
automatic (even in the classical theory), while the presence of
Einstein-like dynamics is almost guaranteed by one-loop quantum
effects.
\section*{Acknowledgments}

The authors wish to thank Sebastiano Sonego for his comments and suggestions.


\clearpage

\end{document}